\documentstyle[12pt]{article}
\begin{document}

\title{Answer to Question 55.\\
Are there pictorial examples that distinguish covariant and contravariant
vectors?}
\author{Bernard Jancewicz\\
Institute of Theoretical Physics, University of Wroc{\l}aw, \\
pl. Maksa Borna 9, PL-50-204 Wroc{\l}aw, Poland}
\maketitle

The pair of terms: {\it contravariant vectors} and {\it covariant vectors}
can be replaced by another pair: vectors and linear forms. The vectors 
constitute the linear space, that is a set with two operations: addition and 
multiplication by scalars. The linear form is a mapping of vectors into 
scalars which is additive and homogeneous under multiplication by scalars. 
The set of linear forms is itself a linear space -- it is called {\it dual 
space}.  Neuenschwander [1] claims that "the disctinction between covariant 
and cotravariant vectors is necessary when dealing with curved spaces." I 
would add that the distinction between vectors and linear forms is necessary 
also in a vector space\footnote{Strictly speaking, we should say "in an
affine space". This remark is for reader with more rigorous mathematical
knowledge.} devoid of scalar product and metric determined by it. As was shown
in [2], the concept of flat space without metric is useful in describing
electrostatics in
anisotropic dielectric medium and magnetostatics in anistropic magnetic 
medium, where various metrics can be introduced, determined by the 
permittivity or permeability tensors.

One should be aware of the fact that not all familiar geometric notions exist
in\break a vector space devoid of the scalar product. Valid notions include:
linearly dependent vectors, linearly independent vectors, parallel vectors and
parallel planes. Undefined notions include: angles, perpendicular vectors,
comparable length of nonparallel vectors, circles and spheres.

We depict vector as a directed segment. Its relevant features are direction
and magnitude. The {\it direction} consists of a straight line (on which the
vector lies) which, after Lounesto [3], can be called an {\it attitude} of
the vector, and an arrow on that line which we call an {\it orientation}. Two
vectors of the same attitude are parallel.

In the case of a vector space without a scalar product, there is a need to
introduce dual objects known as {\it linear forms} which, in a sense, replace
the scalar product. 
The {\it linear form} (or {\it one-form} or {\it outer form of first order})
is a linear mapping of vectors into scalars: ${\bf f}:~{\bf x}\to {\bf f}(%
{\bf x})\in {\bf R}$. It is known from algebra that its {\it kernel}, i.e.
the set $M=\{{\bf x}:~{\bf f}({\bf x})=0\}$ is a linear subspace of ${\bf 
R}^3$ with dimension two, so $M$ is a plane. Thus, for each linear form, one 
can associate a plane $M$ passing through the origin. One can also find other
planes parallel to $M$, on which the form ${\bf f}$ takes values 1,~2,~3 and
so on. In this way, we get the geometric image of a form as a family of
equidistant parallel planes with an arrow joining the neighbouring planes
and showing the direction of increase of the form; see Figure 1.
The angle between the arrow and the planes is not important.

Conversely, once we have such a family of planes, a number can be ascribed to 
any vector ${\bf x}$ as follows. If the origin of ${\bf x}$ lies on one of the
planes, we count the number of planes pierced by ${\bf x}$ and this is equal
to ${\bf f}({\bf x})$; see Figure 1. If ${\bf x}$  ends on some plane, the
number ${\bf f}({\bf x})$ is integer; otherwise ${\bf f}({\bf x})$ 
is not integer. If ${\bf x}$ intersects the planes in the direction of 
decreasing labels of planes, then ${\bf f}({\bf x})$ is negative. This 
prescription, along with the corresponding pictures, can be found in 
Ref.~[4]. Similar way of visualisation of the linear forms is in the
books of Burke [5,6].

\begin{center}
\begin{picture}(170,100)
\put(0,5){\line(1,0){120}}
\put(0,5){\line(3,2){30}}
\put(18,8){\it 0}
\put(0,25){\line(1,0){120}}
\put(0,25){\line(3,2){30}}
\put(18,28){\it 1}
\put(0,45){\line(1,0){120}}
\put(0,45){\line(3,2){30}}
\put(18,48){\it 2}
\put(0,65){\line(1,0){120}}
\put(0,65){\line(3,2){50}}
\put(18,68){\it 3}
\put(50,98){\line(1,0){120}}
\put(120,5){\line(3,2){50}}
\put(120,25){\line(3,2){50}}
\put(120,45){\line(3,2){50}}
\put(120,65){\line(3,2){50}}
\put(140,78){\line(1,0){29}}
\put(140,58){\line(1,0){29}}
\put(140,38){\line(1,0){29}}
\put(0,9){\vector(0,1){12}}
\put(45,15){\thicklines \line(1,1){10}}
\put(65,35){\thicklines \line(1,1){10}}
\put(85,55){\thicklines \line(1,1){10}}
\put(105,75){\thicklines \vector(1,1){10}}
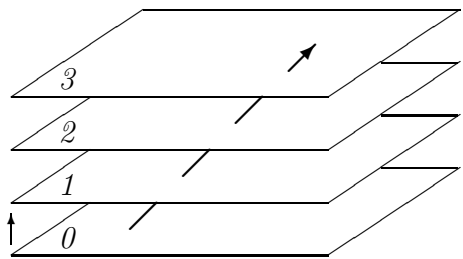
\put(-30,-20){Figure 1. Geometric image of a linear form.}
\end{picture}
\end{center}

\vspace{7mm}
As a matter of fact, in order to describe a linear form one does not need to 
draw infinitely many parallel equidistant planes. Two neighbouring ones are
sufficient; see Figure 2. Thus we claim that the geometric image of a
linear form is a {\it slab} (a {\it plane-parallel layer)} with an arrow 
penetrating it from one boundary to the other. Pictures of this kind were
shown already in Ref. [7].
\begin{center}
\begin{picture}(150,60)
\put(0,5){\line(1,0){120}} 
\put(0,25){\line(1,0){120}}
\put(0,25){\line(3,2){50}}
\put(50,58){\line(1,0){120}}
\put(120,5){\line(3,2){50}}
\put(120,25){\line(3,2){50}}
\put(50,9){\vector(1,2){6}}
\put(0,5){\line(0,1){20}}
\put(170,38){\line(0,1){20}}
\put(120,5){\line(0,1){20}}
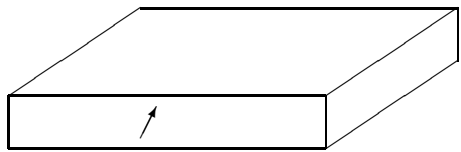
\put(-50,-20){Figure 2. Simplified geometric image of a linear form.}
\end{picture}
\end{center}

\vspace{7mm}
\noindent In this manner, we arrive to conclusion that direction of the 
linear form consists of {\it attitude} -- a plane and {\it orientation} -- 
an arrow piercing the plane. There are also pictorial prescriptions
of adding linear forms and multiplying them by scalars, we refer the reader
to Refs. [2] and [6].

Strictly speaking, in mathematics, the duality notion is reversible:
when linear space $V^{\prime }$ is dual to a linear space $V$, then $V$
is also dual to $V^{\prime }$. Hence one could ask the question:
Why do we visualise elements of $V$ as directed segments, and elements
of $V^{\prime }$ as families of parallel planes? Why not vice versa?
My answer is as follows: We all live in the three-dimensional where the
position vector {\bf r}, called also radius vector, is naturally 
represented by the directed segment. Therefore, if we put {\bf r} in $V$,
all elements of $V$ should be represented also by segments and,
consequently, all elements of $V^{\prime }$ by families of planes.
When one puts {\bf r} in $V^{\prime }$, then everything is other way
around.

It is useful to give some examples of physical quantites which can be
regarded undoubtedly as vectors and others as linear forms. The most natural
vectorial quantity is the {\it radius vector} {\bf r} of a point in space
relative to the reference point, called the {\it origin of a frame}. The {\it
displacement vector} {\bf l} is of the same nature. Hence the {\it velocity}
${\bf v}=d{\bf r}/dt$, the derivative of {\bf r} with respect to a scalar
variable $t$, is also a vector. The same is true of the {\it acceleration}
${\bf a}=d{\bf v}/dt$ and the {\it electric dipole moment} ${\bf d}=q{\bf l}$.

If one considers the potential energy $U$ as a scalar, its relation to force
in the
traditional language is $dU=-{\bf F}\cdot {\bf dr}$ where dot denotes the
scalar product. This means that force is a linear map of the infinitesimal
vector $d{\bf r}$ into the infinitesimal scalar $-dU={\bf F}(d {\bf r})$. 
Thus, after this observation, the force should be treated as a linear form. 
This in turn, by the Newton equation ${\bf F}=d{\bf p}/dt$, implies that the 
momentum {\bf p} has to be a linear form too. This view is adopted in modern
mathematical formulations of mechanics, see e.g. [8].

A one-form quantity occurs also in the description of the plane waves. The
locus of points in space with the same phase of a wave is just a plane. The
family of planes with phases differing by $2\pi n$ for natural $n$ can be
viewed as the geometric image of a one-form depicted in Figure 1.
This one-form describes the physical quantity known as the {\it wave vector} 
{\bf k} with magnitude $2\pi /\lambda $ ($\lambda $ is the wavelength).
Thus, in my opinion, the physical quantity {\bf k} should have a
different name since, in its directed nature, it is not a vector. If 
I may create an English word I would propose the name {\it wavity} 
for~{\bf k}.

Another one-form is the {\it electric field strength} {\bf E}, since we
consider it to be a linear map of the infinitesimal vector $d{\bf r}$ into
the infinitesimal potential difference: $-dV={\bf E}(d{\bf r})$.
The attitute of {\bf E} at each point of space is plane tangent to
the equipotential surface passing through the point.

It is worth mentioning that there is a plenty of other geometric objects
which can be called {\it directed quantities} in the sense that the
direction and magnitude can be ascribed to them. Among them
primary directed quantities are {\it multivectors:} vectors,
bivectors, and trivectors. Just as a vector, in the process of abstraction,
arises from a straight line segment with an orientation, a {\it bivector}
originates from a plane segment with an orientation, and a {\it trivector}
from a solid body with an orientation. The connection of multivectors to
straight lines, planes and bodies gives them the advantage of being easily
depicted in illustrations. For their application in classical mechanics and 
in electromagnetism, see Ref. [9] and [10] respectively.

{\it Outer forms} are quantities dual to multivectors. They are called {\it 
differential forms} when they depend on their position in space. Differential 
forms are very popular in theoretical physics, see eg. Refs.~[11-13], but
authors writing about them rarely use pictures to illustrate the concept to 
the reader; nice exceptions are already mentioned Refs.~[4--6] in which 
outer forms are presented as specific ``slicers''. No care is put there on 
directions of outer forms. Careful description of directions of the whole 
variety of directed quantities can be found in Refs. [2] and [14].

\vspace{5mm}
{\bf Acknowledgement}

\medskip
I am grateful to Andrzej Borowiec who turned my attention to this question.

\end{document}